\documentstyle[12pt,epsfig]{article}

\newcommand{\beq}{\begin{equation}}
\newcommand{\eeq}{\end{equation}}
\newcommand{\bea}{\begin{eqnarray}}
\newcommand{\eea}{\end{eqnarray}}

\newcommand{\gsim}{\lower.7ex\hbox{$
\;\stackrel{\textstyle>}{\sim}\;$}}
\newcommand{\lsim}{\lower.7ex\hbox{$
\;\stackrel{\textstyle<}{\sim}\;$}}

\def\lsim{\mathrel{\rlap{\lower3pt\hbox{\hskip0pt$\sim$}}
    \raise1pt\hbox{$<$}}}         
\def\gsim{\mathrel{\rlap{\lower4pt\hbox{\hskip1pt$\sim$}}
    \raise1pt\hbox{$>$}}}         

\renewcommand{\Im}{{\rm Im}\,}

\newcommand{\bibit}[1]{\bibitem{#1}}

\newcommand{\aver}[1]{\langle #1\rangle}

\newcommand{\La}{\overline{\Lambda}}

\newcommand{\Lam}{\Lambda_{\rm QCD}}
\newcommand{\mhad}{\mu_{\rm hadr}}

\newcommand{\tto}{\!\to\!}

\newcommand{\MeV}{\,\mbox{MeV}}
\newcommand{\matel}[3]{\langle #1|#2|#3\rangle}

\newcommand{\msp}[1]{\mbox{\hspace*{#1mm}~}}

\begin{document}
\thispagestyle{empty}
\vspace*{-10mm}

\begin{flushright}
Bicocca-FT-04-10\\
UND-HEP-04-BIG\hspace*{.08em}06\\
hep-ph/0407359
\end{flushright}
\vspace*{12mm}

\begin{center}
{\LARGE{\bf
On the strong-coupling effects in jet \vspace*{4mm} \\
bremsstrahlung in the heavy flavor decays}}
\vspace*{15mm}

\end{center}

\begin{center}
{\LARGE Nikolai~Uraltsev}\hspace*{1.5pt}\raisebox{5.5pt}{*}
\vspace*{8mm} \\
{\sl INFN, Sezione di Milano,  Milano, Italy}\vspace*{.5mm}\\
{\small {\sf and}}\vspace*{.5mm} \\
{\sl Department of Physics, University of Notre Dame du Lac,
Notre Dame, IN 46556, USA}
\vspace*{40mm}

{\bf Abstract}\vspace*{-.9mm}\\
\end{center}

\noindent
We address the strong-coupling regime effects in the
light-like quark jets where the radiated gluons are hard but
highly collinear. These may lead to additional
contributions in the invariant mass (or recoil energy)
spectra, on top of the Fermi motion of the initial heavy
quark in decays like $b\tto s+\gamma$. It is shown
that the integer moments are nevertheless
free from such effects; perturbation theory corrections for the
moments are driven by $\alpha_s$ in the small coupling
regime. They are modified nonperturbatively by the soft modes but not
by the collinear modes.

\setcounter{page}{0}

\vfill

~\hspace*{-12.5mm}\hrulefill \hspace*{-1.2mm} \\
\footnotesize{
\hspace*{-5mm}$^*$On leave of absence from
St.\,Petersburg Nuclear Physics
Institute, Gatchina, St.\,Petersburg  188300, Russia}
\normalsize

\newpage

\section{Introduction}

The OPE provides a consistent approach to describe
sufficiently inclusive distributions in the decays of
heavy flavor hadrons. It incorporates power-suppressed effects
which originate from the nonperturbative domain that is
responsible for
confinement and bound-state dynamics. Its important
ingredient is the separation between the contributions of different
momentum scales, with the short distances generically
referred to the coefficient functions, and large distances
$\sim \Lam^{-1}$ belonging to the nonperturbative
operators. This general approach applies to both the weak
transitions into sufficiently heavy quarks like $b\tto
c\,\ell\nu$, and to heavy-to-light decays
like $b\tto s+\gamma\,$ or $\,b\tto u \,\ell\nu$. The former
case has been best elaborated including the technical
implementation of the Wilsonian `hard' momentum scale
separation in the perturbative expansion. This approach
appeared successful in describing experimental data
\cite{babarprl}. There remain, however certain principal
differences between the decays into heavy and light-like
quarks when gluon bremsstrahlung is carefully incorporated.

The standard application of the OPE is initially
formulated in the Euclidean theory. It utilizes the expansion
of the amplitudes in the momentum of the soft quantum
fields which describe nonperturbative dynamics. Therefore, it
assumes all the components of the momentum to be bounded by
some mass $\mu$ identified with the normalization
scale of the nonperturbative operators. It is normally
chosen to be
$$
\mu\simeq \mbox{~few} \times \mhad\,,
$$
where $\mhad\!\simeq\! 600\MeV$ represents the typical momentum
scale characteristic for the strong coupling domain in
QCD. Consequently, for the gluon exchanges those with {\sf any}
component of the momentum $k_\alpha$ larger than $\mu$ are
attributed to the coefficient functions and are calculated in practice
in perturbation theory. The same idea is transferred
with minimal modifications to decays in the Minkowski
kinematics.

This procedure works fine for the transitions between heavy
quarks, in particular in $b\tto c\,\ell\nu$. The lower
cutoff on the gluon energy automatically eliminates the infrared
domain and no running coupling $\alpha_s(q^2)$
enters at the scales $q^2 \!\ll\! \mu^2$. The perturbative
expansion is therefore applicable and justified.

The situation looks different in the heavy-to-light decays, in
particular in $b\tto s+\gamma$. The gluon bremsstrahlung is
described by the double logarithmic probability
\beq
{\rm d} W= \int \!\frac{{\rm d}\omega}{\omega}\;
\int \!\frac{{\rm d}k_\perp^2}{k_\perp^2}\:
C_F \frac{\alpha_s(k_\perp^2)}{\pi} \;\,{\rm d} W_{\rm born}\;,
\label{20}
\eeq
where ${\rm d} W_{\rm born}$ is the bare `hard' decay rate and
$k_\mu\!=\!(\omega, k_\perp, k_{{^\|}})\,$ is the gluon
four-momentum. This expression literally assumes $\omega \!\ll\!
m_b$, $k_\perp \!\ll\! m_b$ and the radiation angle
$\theta\!\simeq\! \frac{k_{{^\|}}}{\omega}\!\ll\! 1$. More
accurate expressions for arbitrary $\theta$ do not change
the principal point here.

The radiation probability (\ref{20}) underlies the conceptual
problem: even if the gluon is very energetic by itself,
$\omega \!\gg\! \mu$, the transverse momentum $k_\perp$ can be
low, $|k_\perp|\!\lsim\! \Lam$ if the gluon is highly
collinear with $\theta \!\ll\! \frac{\mu}{\omega} \!<\!1$. The
`perturbative' bremsstrahlung then runs into the
strong-coupling domain and cannot be unambiguously evaluated
expanding in $\alpha_s(Q^2)$ with $Q^2\! \gsim \mu^2$. This
observation may cast doubts on calculability of even the
`robust' observables like the moments of the decay distributions.

A more careful analysis, however suggests that the primary
OPE results for the (integer) moments relating them to the
local heavy quark operators, remain unchanged, while the
spectra themselves, in general, may possibly
depend to some extent on new strong-coupling effects
appearing in jet physics.

\section{OPE for the decay distributions in \boldmath $b\tto q$}

For the sake of simplicity we will phrase the subsequent
discussion for the decay $b\tto s+\gamma$ (or, generically,
$Q\tto q+\varphi$ with $m_q\!=\!0$ and $m_\varphi\!=\!0$,
referring to the colorless $\varphi$ as a photon whether or
not it carries spin). We will also assume a much stronger
hierarchy $m_b\!=\!m_Q \gg \mhad$ than exists in reality, so that we
may discard inessential $1/m_b$ corrections ab initio.

The photon spectrum in $b\tto s+\gamma$ is represented by
the convolution of the soft `primordial' distribution
function $F(k_+)$ often known \cite{prl,motion} as
describing ``Fermi motion''
\cite{ap,acm} of the $b$ quark inside $B$ meson, and of the
`hard' spectrum evolving from the bare two-body
$\delta(E_\gamma\!-\!\frac{m_b}{2})$ due to the gluon
bremsstrahlung:
\beq
\frac{{\rm d}\Gamma_{\rm tot}(E_\gamma)}{{\rm d}E_\gamma}=
\int_{-\infty}^\infty \!{\rm d}k_+\: F(k_+)\,
\frac{{\rm d}\Gamma_{\rm pert}}{{\rm d}E}
(E_\gamma\!-\!\mbox{$\frac{k_+}{2}$})\;.
\label{24}
\eeq
The support of $F(k_+)$ actually lies below $\La\!\simeq\!
M_B\!-\!m_b(\mu)$; at
large negative arguments with $|k_+|\!\gg\! \mhad $  the
distribution function $F(k_+)$ must decrease exponentially.

Kinematically we have
\beq
E_\gamma = \frac{M_B^2-M_X^2}{2M_B}\;,
\label{26}
\eeq
where $M_X^2$ is the invariant hadronic mass squared in the
final state (jet invariant mass, in the perturbative
description). Instead of the photon spectrum we can,
therefore speak of the distribution in $M_X^2$, or of the moments
of the hadronic mass squared which has more universal
application in jet physics. The typical nonperturbative
(bound-state) domain in $M_X^2$ (referred to as `window' in
Ref.~\cite{motion}) is $M_X^2 \!\lsim\! \La m_b$, and larger
$M_X^2$ emerges due to hard bremsstrahlung.

As schematically illustrated in Fig.~1, $F(k_+)$ describes
the intrinsic properties of the decaying bound state. It is
universal with respect to any concrete type of $Q\tto q$
transition as long as the recoiling system is colorless and the
energy of the light quark is large. All field modes included
in it have wavelengths limited by $\mu$. Therefore, it is
completely independent of $m_b$ once power corrections
$\sim\!\mhad/m_b$ are neglected.

\thispagestyle{plain}
\begin{figure}[hhh]\vspace*{-3.4mm}
\begin{center}
\mbox{\epsfig{file=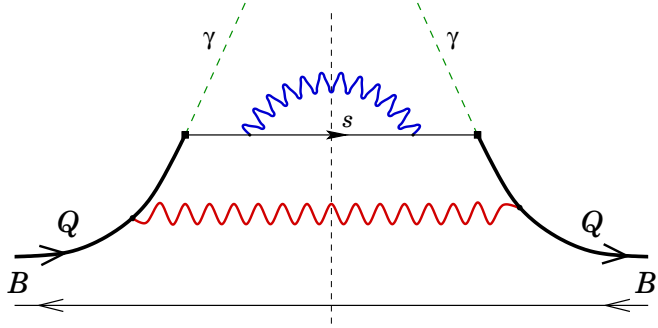,width=90mm
}}
\end{center}
\vspace*{-6.0mm}
\caption{ \small
The corrections to the inclusive decay
distribution in the OPE illustrated with the  perturbative
diagrams. The red gluon line shows the `soft'
modes with all the
components of the four-momentum small. They are
responsible for the `Fermi motion' and are associated with
the decaying bound state. The blue gluon line stands for harder
gluons with $\omega>\mu$ attributed to $\Gamma_{\rm pert}$,
which nevertheless can be highly
collinear and emitted with large coupling.
}
\end{figure}

$F(k_+)$ is constrained by its (integer) moments like
the usual DIS structure functions. The moments are given by the
expectation values of the local heavy quark operators over
the initial hadron:  
\bea
\nonumber
F_0 \msp{-4}&=&\msp{-4}\int \! {\rm d}k_+ \;F(k_+) =1\;, \msp{20}
F_1 =\int \! {\rm d}k_+ \;k_+\,F(k_+)=0\;, \\
\nonumber
F_2 \msp{-4}&=&\msp{-4}\int \!
{\rm d}k_+ \;k_+^2\,F(k_+)=\frac{\mu_\pi^2}{3}\;, \msp{11.3}
F_3 =\int \! {\rm d}k_+
\;k_+^3\,F(k_+)=-\frac{\rho_D^3}{3}\;, \\
F_n \msp{-4}&=&\msp{-4}\int \! {\rm d}k_+
\;k_+^n\,F(k_+)= \frac{1}{2M_{H_Q}\!}
\matel{H_Q}{\bar{Q}iD_z(iD_0\!-\!iD_z)^{n-2}iD_z Q}{H_Q}\;.
\label{28}
\eea

In Eq.~(\ref{24})
$\frac{{\rm d}\Gamma_{\rm pert}}{{\rm d}M_X^2}$ accounts for
all other gluon modes. In practical terms, the renormalization
scheme \cite{ren} is advantageous where, at the level of one
gluon emissions the normalization procedure sets the separation
based on the gluon energy $\omega$. Then, for radiating a
single gluon we have to perform the integration with only
the lower cutoff on $\omega$, for instance
\beq
\frac{{\rm d}\Gamma^{\rm pert}}{{\rm d}M_X^2}=
\int \!\frac{{\rm d}\omega}{\omega}\;\vartheta(\omega\!-\!\mu)
\int \!\frac{{\rm d}k_\perp^2}{k_\perp^2}\;
C_F\frac{\alpha_s(k_\perp^2)}{\pi}
\;\delta(M_X^2\!-\!k_\perp^2 \mbox{$\frac{m_b}{2\omega}$}) \;.
\label{30}
\eeq
Here $m_b/2$ simply stands for the initial energy of the
$s$ quark jet $E_{\rm jet}$.
The kinematic enhancement factor $m_b/(2\omega)$ in $M_X^2$
brings in the contribution of large transverse distances
$x_\perp \!>\! \mhad^{-1}$ into the calculation of the perturbative
spectrum in Eq.~(\ref{30}) once $M_X^2$ becomes smaller than
$\frac{m_b}{2}\mhad \!\cdot\! \frac{\mhad}{\mu}$. The factor
$\frac{\mhad}{\mu}$ is smaller than unity, yet it represents
a numeric rather than parametric suppression.
Perturbation theory alone is unable to
precisely calculate its decay spectrum closer to
the endpoint than a fraction of the distribution function
domain (window), $\mhad \!\cdot m_b$. This fraction
appears to depend
on the normalization point. Consequently, the full point-to-point
spectrum may not be completely expressed through the
distribution function with the resolution arbitrarily higher
than the size of the Fermi motion `window'.

It is important that the type of the strong-coupling effects
generated by the `perturbative' jet function Eq.~(\ref{30})
is physically distinct from the Fermi motion. The latter is
determined by the bound-state dynamics and depends on the
$B$ meson wavefunction at distances $\sim \mhad^{-1}$
from the heavy quark and its decay vertex. The jet
hadronization, on the contrary, runs into the domain of
the nonperturbative coupling only over the time interval $\sim
\mhad^{-1}$ in the frame accompanying the final light
quark, a much longer period $\tau_{\rm jet} \gsim
\mhad^{-1} \sqrt{\frac{m_b}{\mhad}}$due to the Lorentz
slowdown. The hadronizing system travels over a distance
$\gsim m_b^{\frac{1}{2}}\mhad^{-\frac{3}{2}}$ before
that. The nonperturbative hadronization occurs
far away in space from the decay point where the original
$B$ meson was located. This suggest that the considered
nonperturbative jet effects are largely universal and are
independent of the properties of the initial bound
state. There are arguments that this separation is
exponential in the ratio
$\mu/\mhad$.\footnote{Yu.~Dokshitzer, private communication.}
We will return to this point later.

\section{Moments of the distribution}

The incalculable nonperturbative end point in $\frac{{\rm d}
\Gamma^{\rm pert}}{{\rm d}M_X^2}$ and the associated
indeterminacy of $\frac{{\rm d} \Gamma_{\rm tot}}{{\rm d}M_X^2}\,$
raise the concern of whether the moments of the total
spectrum can still be obtained, in the presence of gluon
bremsstrahlung, in terms of the local heavy quark
expectation values applying only the truly short-distance corrections
evaluated in perturbation theory over the
small-coupling domain. At first glance, Eq.~(\ref{30}) mandates
the presence of additional nonperturbative corrections from jet
physics at small $k_\perp^2$, at least at the level
$(m_b\mhad)^n\!\cdot\!(\mhad/\mu)^n$ for the $n$th
moment. In the zeroth moment of $\frac{{\rm d}
\Gamma^{\rm pert}}{{\rm d}M_X^2}$, the total decay rate
$\Gamma_{\rm pert}$ such effects cancel between the
bremsstrahlung and the virtual corrections as required by the
KLN theorem. The virtual effects are absent from the
positive moments, and the simple-minded application of
Eq.~(\ref{30}) yields non-vanishing nonperturbative
contributions already to $\aver{M_X^2}$ proportional to the
first moment ${\cal A}_2\!=\!-A_2$ of the
effective coupling (the $\log$-moment of the dispersion
coupling $\alpha_{\rm eff}(s)$) \cite{dmw2}
parameterizing power corrections in a number of observables
in jet physics.

We find, however that such a conclusion would be
incorrect. The nonperturbative jet effects, although
possibly present
in the differential distribution itself, cancel out in the
integer moments. The bremsstrahlung corrections to the usual
nonperturbative expressions for the moments in terms of the
$B$-meson local heavy quark expectation values, calculated in
perturbation theory, do not involve the running coupling at
scales below $\mu$.\footnote{We do not consider the moment
rank $n$ as a large parameter and, therefore do not
distinguish between, say $\alpha_s(\mu)$ and $\alpha_s(\mu/\sqrt{n})$.}

The most transparent way to illustrate physics behind the
related `conspiracy' in the strong-coupling jet effects
uses the so-called dispersive approach to
perturbation theory in Minkowski space scrutinized by
Dokshitzer, Marchesini and Webber \cite{dmw}; a recent brief
summary for purely perturbative BLM-type applications can be
found in Refs.~\cite{blmvcb}. The method is based on the
dispersion representation for the dressed gluon propagator
\beq
\frac{\alpha_s^E(Q^2)}{Q^2} = \pi \int \! \frac{{\rm d}
\lambda^2}{\lambda^2}
\;\frac{\rho(\lambda^2)}{\lambda^2+Q^2}\;, \qquad
\rho(s) = -\frac{1}{\pi^2} \Im\alpha_s(-s)
\label{40}
\eeq
via a fictitious gluon mass $\lambda$; $Q^2$ is the
Euclidean virtuality and $\alpha_s^E$ denotes the standard
Euclidean running coupling
(the above expression for simplicity assumes no pole in
$\alpha_s^E(Q^2)/Q^2$ at $Q^2\!=\!0$).
This requires calculating the
inclusive observable in question $O$ to order $\alpha_s$
with an arbitrary gluon mass $\lambda$,
$O^{\rm born} \!+\! C_{F\,}\frac{\alpha_s}{\pi}\,O_1(\lambda^2)$.
The effect of the running
$\alpha_s$ is then obtained by
\beq
O^{\rm resum}= O^{\rm born}+C_F
\int \! \frac{{\rm d}
\lambda^2}{\lambda^2}
\;\rho(\lambda^2)\,O_1(\lambda^2)
\label{42}
\eeq
(the order-$\beta_0\alpha_s^2$ version of this BLM improvement
was formulated in Ref.~\cite{volblm}).
In this context the OPE separates the effects into
`perturbative' and `nonperturbative' according to their
behavior in the deep Euclidean domain, not for the Minkowski
objects. Therefore, one puts
\beq
\rho(\lambda^2)=\rho_{\rm pert}(\lambda^2)+\delta\rho(\lambda^2)
\label{44}
\eeq
and it is assumed for the corresponding dispersion
integrals, $\alpha_s^E(Q^2)$ and $\delta\alpha_s^E(Q^2)$,
that the latter dies out fast at large $Q^2>0$.

Being interested in the effects from the strong-coupling
regime we need to retain only the $\delta\alpha_s^E(Q^2)$,
or $\delta\rho(\lambda^2)$ piece. For simplicity we will use
the full coupling in the following equations, however,
implying that the `nonperturbative' part can be separately
considered when required.

The standard bremsstrahlung probability within this
framework takes the following form:
\beq
{\rm d}W_{\rm brem}=  C_F \int \!\frac{{\rm d}\omega}{\omega} \;
\int \! \frac{{\rm d}\lambda^2}{\lambda^2}\;\rho(\lambda^2)
\int \!\frac{{\rm d}k_\perp^2}{k_\perp^2+\lambda^2}\;
{\rm d} W_{\rm born}\;.
\label{46}
\eeq
The integral over $\lambda^2$ is nothing but the dispersion
representation for $\frac{\alpha_s(k_\perp^2)}{k_\perp^2}$:
\beq
\int \! \frac{{\rm d}\lambda^2}{\lambda^2}\;\rho(\lambda^2)
\frac{1}{k_\perp^2+\lambda^2}= \frac{1}{\pi}\,
\frac{\alpha_s(k_\perp^2)}{k_\perp^2}\;,
\label{48}
\eeq
justifying the standard prescription Eq.~(\ref{20}).

As already mentioned, there is the KLN cancellation of soft
gluons between the bremsstrahlung and the virtual
corrections. The latter do not contribute to
the higher moments, however we get generally
the power-divergent integrals like
$\int \! \frac{{\rm d}k_\perp^2}{k_\perp^2} (k_\perp^2)^n
\alpha_s(k_\perp^2)$. The effective $\alpha_s(k_\perp^2)$
then has to be accounted for with higher accuracy. To do
this we write explicitly, e.g.\ for the first moment
\beq
\aver{M_X^2}^{\rm pert}=C_F \int \!{\rm d}M_X^2 \int \!
\frac{{\rm d}\omega}{\omega}\, \vartheta(\omega\!-\!\mu)
\int \! \frac{{\rm d}\lambda^2}{\lambda^2}\;\rho(\lambda^2)
\int \!\frac{{\rm d}k_\perp^2}{k_\perp^2+\lambda^2}\;
M_X^2
\delta(M_X^2\!-\!(k_\perp^2\!+\!\lambda^2)\mbox{$\frac{m_b}{2\omega}$})
 \;.
\label{50}
\eeq
Performing the integration over $M_X^2$ yields
\beq
\aver{M_X^2}^{\rm pert}= C_F \int \!
\frac{{\rm d}\omega}{\omega}\, \vartheta(\omega\!-\!\mu)\;
\frac{m_b}{2\omega}\;
\int \! \frac{{\rm d}\lambda^2}{\lambda^2}\;\rho(\lambda^2)\;
\int \!{\rm d}k_\perp^2 \;.
\label{52}
\eeq
The integral has split into the product of independent integrals
over $k_\perp^2$ and over $\lambda^2$, which means that the
effective coupling enters only at the high scale $\sim\!
\sqrt{\mu\, m_b}$ or higher, cf.\
Eqs.~(\ref{60}), (\ref{64}), determined
by the effective ultraviolet cutoff in the
integral. Formally, the absence of the nonperturbative
contributions follows from
\beq
\pi \int \! \frac{{\rm d}\lambda^2}{\lambda^2}\;
\delta\rho(\lambda^2)=
\lim_{Q^2 \to \infty} \delta \alpha_s^E(Q^2)=0
 \;.
\label{54}
\eeq

The expression for the spectrum itself helps to interpret
this general result:
\bea
\nonumber
\frac{{\rm d}\Gamma^{\rm pert}}{{\rm d}M_X^2} \msp{-4}&=&\msp{-4}
C_F \int \!\frac{{\rm d}\omega}{\omega}\;\vartheta(\omega\!-\!\mu)\;
\int \! \frac{{\rm d}\lambda^2}{\lambda^2}\:\rho(\lambda^2)\;
\int \!\frac{{\rm d}k_\perp^2}{k_\perp^2\!+\!\lambda^2}\;
\delta(M_X^2\!-\!(k_\perp^2\!+\!\lambda^2) \mbox{$\frac{m_b}{2\omega}$})\\
\msp{-4}&=&\msp{-4}
\frac{C_F}{M_X^2}
\int \!\frac{{\rm d}\omega}{\omega}\;\vartheta(\omega\!-\!\mu)\;
\int \!
\frac{{\rm d}\lambda^2}{\lambda^2}\:\rho(\lambda^2)
\;\vartheta(M_X^2\!-\!\mbox{$\frac{m_b}{2\omega}$}\lambda^2)
\;.
\label{60}
\eea
This differs from the expression
\beq
\frac{C_F}{M_X^2}
\int \!\frac{{\rm d}\omega}{\omega}\;\vartheta(\omega\!-\!\mu)\,
\frac{\alpha_s^E(\mbox{$\frac{2\omega}{m_b}$}M_X^2)}{\pi}
\label{62}
\eeq
which would literally follow from the prescription
(\ref{30}). The effective coupling in the spectrum $\tilde
\alpha_s(\mbox{$\frac{2\omega}{m_b}$}M_X^2)$ differs from the
usual Euclidean coupling
$\alpha_s(\mbox{$\frac{2\omega}{m_b}$}M_X^2)$:
\beq
\tilde\alpha_s(Q^2)= \pi \int_0^{Q^2}\!
\frac{{\rm d}\lambda^2}{\lambda^2}\:\rho(\lambda^2)  \qquad
\mbox{while} \qquad
\alpha_s^E(Q^2)= \pi \int_0^{\infty}\!
\frac{{\rm d}\lambda^2}{\lambda^2+Q^2}\:\rho(\lambda^2)
\label{64}
\eeq

The two couplings agree in the perturbative
domain with `$\log$' accuracy where the momenta appear
ordered in their scale. Yet they become different in the strong
coupling regime. In particular, the moments of
$\delta\alpha_s^E(Q^2)$ are nonzero and, for the simplest
ansats\"{e} \cite{dmw} all have the same sign scaling as
$\mhad^{2n}$. The integer moments of $\delta
\tilde\alpha_s(Q^2)$, on the contrary, all
vanish \cite{bsg,dmw} according to
the generalization of the relation (\ref{54}):
\beq
\pi \int \! \frac{{\rm d}\lambda^2}{\lambda^2}\,\lambda^{2n} \;
\delta\rho(\lambda^2)=(-1)^n
\lim_{Q^2 \to \infty} Q^{2n}\,\delta \alpha_s^E(Q^2)
 \;.
\label{68}
\eeq

While the general expression for the probability of the
gluon bremsstrahlung, Eq.~(\ref{20}) remains valid, in a
sense, even nonperturbatively, the actual physical effect
differs from what would follow from Eq.~(\ref{20}) at the
power level. The latter therefore has limited applicability. The
reason is that at this level the one-particle massless-gluon
description of the
interaction becomes incompatible with
running of $\alpha_s$, a field-theory effect associated with
a few particle states of somewhat different
kinematics. Having the gluon split, while not modifying the
total rate, changes fine kinematic details thus reshuffling the
distributions. This modifies the effective coupling for
differential distributions, making it observable-dependent
once the result is forcibly interpreted in terms of the massless
one-gluon framework. Additionally, these effective couplings
obey certain integral constraints which effectively recognize
that the Euclidean image of the inclusive bremsstrahlung is
a genuinely short-distance process.

The result of the calculations supports the interpretation
that, in the hard collinear jet configurations the growth of
$\alpha_s$ from the initial $\alpha_s(E_{\rm jet})$ to
$\alpha_s(k_\perp)$ is an effect of the final-state
interaction, viz.\ jet splitting. As such, it is not
expected to affect fully inclusive truly short-distance
characteristics like the total decay rate or recoil
moments. We see that the long-time jet contributions
disappear in just the integer moments of $M_X^2$, indicating
that $M_X^2$ is the right kinematic variable.

\section{Discussions}

Jet hadronization effects may affect the hadronic mass
distribution and, consequently, the point-to-point recoil
spectrum in the decays like $B\tto X_s+\gamma$ or $B\tto
X_u\,\ell\nu$. Yet it is shown that there remain no such
nonperturbative strong-coupling--domain contributions to the
moments. The perturbative corrections to the OPE relations
for the moments come from the small coupling
regime. Nonperturbatively the latter are shaped by the soft
modes but
not by the collinear modes.

There appears an interesting analogy between the
strong-coupling effects in high-energy jets and local
duality violation in the heavy quark decays specifically
studied in Ref.~\cite{inst} in the instanton vacuum
ansatz. In both cases the nonperturbative effects were
present and enhanced close to the end point in the
differential distributions. Integrating them over the
available kinematic domain might seem to only increase the
effect. Yet for the right moments the integrated effect is
always decreased being driven by the suppression in the
corner of the underintegrated domain; it is minimal for
totally inclusive rates. This similarity, perhaps has its roots
in the general properties of the OPE, although the precise
structure of the latter for jet physics still
remains to be understood.

The identified property of the large-energy low-$k_\perp$ gluon
radiation leading to the absence of the soft domain
contribution from the integer moments, seems to have
something in common
with the observation by Beneke and Braun \cite{bb} of the
vanishing of a certain class of the leading, linear in $\Lam/E$
corrections in the Drell-Yan production.

The literal incalculability of the spectrum itself in the
large-$m_Q$ limit near the end point may still seem
unsatisfactory. It may be that there exists a further
conspiracy in the small-$k_\perp$ jet splitting which
eliminates this. A possible source for insights is to examine
the formal $\mu$-dependence of the incalculable
piece. The spectrum must be $\mu$-independent; at the same
time, the kinematic $M_X^2$-domain affected by
large-$x_\perp$ physics appears to depend on the lower
gluon energy cutoff $\mu$. The properties of the thus defined
$F(k_+;\mu)$ have not been studied in detail, however, which
precludes us from definite conclusions. The problem deserves
further dedicated analysis.

It is conceivable that such nontrivial effects at the
interface of the gluon bremsstrahlung and of the genuine
nonperturbative corrections in the point-to-point spectrum
are related to the specific way one defines the primordial
distribution function assuming the Wilsonian cutoff on all
the components of the momentum of the soft modes. Such
a definition  is
motivated by the physics of the heavy quark bound state, but it
looks foreign to the light cone approach. In the infinite
momentum
frame accompanying the jet the light-cone
$k_\perp \!\leftrightarrow\! x_\perp$ and
$E_+ \!\sim\! M_X^2$ are natural variables. Using them for the
separation of mass scales would allow, however large wave vectors
$k_{^\|}$ which are foreign to the bound-state dynamics.

Including the low-$k_\perp$ high-energy jet hadronization
effects into the definition of an effective distribution
function $F(k_+)$ may seem to be a practical way to get
around the incalculability of the perturbative spectrum. This
idea might seem to be supported by the mentioned universality
of the jet hadronization. The advantage is doubtful,
however. Long-wavelength `Fermi motion' and highly
collinear jet interactions seem to describe quite distinct
physics, as mentioned in Sect.~2. Likewise the jet
effects have little in common with the $B$
meson expectation values of the local
heavy quark operators determining the
moments of $F(k_+)$, or with the heavy quark bound-state
dynamics in general. Moreover, they would a priori bring in
the intrinsic (logarithmic) dependence of $F(k_+)$ on the
high scale $m_Q$, from which the conventional definition
is free.

The strong-coupling regime effects addressed above are not
specific to the weak decays of heavy flavors, but are
common to general high-energy light-like jet processes in
hadronic physics. We specifically phrased the discussion in
the context of the
inclusive heavy quark decays since here it can be put in a
more rigorous context of the OPE, absorbing the soft modes
into the heavy quark distribution function $F(k_+;\mu)$.

In our analysis no indications were found towards the specific
additional intermediate momentum scale $\sim \!\Lam m_b$ for
the {\sf inclusive} decay distributions.
The bremsstrahlung integration runs over {\sf all}
virtualities: the concrete domain contributing is simply
determined by the energy of the radiated gluon, ranging from
small virtualities up to $m_b^2$.
\vspace*{4mm}

\noindent
{\bf Acknowledgments:} I greatly benefited from many conversations
with Yu.~Dokshitzer on various aspects of jet physics. It is
a pleasure to thank G.~Marchesini, M.~Shifman, V.~Petrov and
A.~Vainshtein for important discussions and comments. I am grateful to
I.~Bigi and D.~Benson for collaboration on the closely
related aspects of $B$ decays, which initiated the reported
study, and to Th.~Mannel for elucidating discussions of the alternative
approaches.
This work was supported in part by the NSF under grant number
PHY-0087419.

\end{document}